\documentclass[aps,prab,twocolumn,superscriptaddress]{revtex4-1}

\usepackage{amssymb}
\usepackage{lipsum}
\usepackage{amssymb}
\usepackage{bbold}
\usepackage{amssymb}
\usepackage{graphicx}
\usepackage{graphics}
\usepackage{amsmath}
\usepackage{placeins}
\usepackage{hyperref}
\usepackage[dvipsnames]{xcolor}
\usepackage{wasysym}
\usepackage{soul}
\usepackage{float}
\usepackage{array}
\usepackage{tabulary}
\usepackage{caption}
\usepackage{fancyhdr}
\usepackage{wrapfig}

\usepackage[normalem]{ulem}
\usepackage{graphicx}
\usepackage{color,soul}
\usepackage{subfigure}
\usepackage{dcolumn}
\usepackage{bm}
\usepackage{hyperref}
\usepackage{url}
\usepackage{multirow}
\hypersetup{
    colorlinks=true,
    linkcolor=blue,
    filecolor=blue,      
    urlcolor=blue,
    citecolor=blue,
}

\usepackage[makeroom]{cancel}

\begin{document}

\title{Formation of field-induced breakdown precursors on metallic electrode surfaces}

    \author{\firstname{Soumendu} \surname{Bagchi}}
    \affiliation{Theoretical Division, Los Alamos National Laboratory, Los Alamos, NM 87545, USA}
    \author{\firstname{Evgenya} \surname{Simakov}}
    \affiliation{Accelerator Operations and Technology Division, Los Alamos National Laboratory, Los Alamos, NM 87545, USA}
    \author{\firstname{Danny} \surname{Perez}}
    \affiliation{Theoretical Division, Los Alamos National Laboratory, Los Alamos, NM 87545, USA}

\begin{abstract}

Understanding the underlying factors responsible for higher-than-anticipated local field enhancements that trigger vacuum breakdown on pristine metal surfaces is crucial for the development of devices capable of withstanding intense operational fields. In this study, we investigate the behavior of nominally flat copper electrode surfaces exposed to electric fields of hundreds of MV/m. Our novel approach considers curvature-driven diffusion processes  to elucidate the formation of sharp breakdown precursors.
To do so, we develop a mesoscale finite element model that accounts for driving forces arising from both electrostatic and surface-tension-induced contributions to the free energy. Our findings reveal a dual influence: surface tension tends to mitigate local curvature, while the electric field drives mass transport toward regions of high local field density. This phenomenon triggers the growth of sharper protrusions, ultimately leading to a rapid enhancement of local fields and, consequently, system instability.
Furthermore, we delineate supercritical and subcritical regimes across a range of initial surface roughness. Our numerical results align closely with experimentally reported data, predicting critical precursor formation fields in the range of 200 MV/m to 500 MV/m.

\end{abstract}

\maketitle

\section{Introduction}\label{S:intro}
Strong electric fields are often encountered on solid material surfaces of high voltage devices used in a broad range of applications, including high power electronics, compact linear accelerators \cite{sicking2020precision, simakov2018advances}, as well as field-emission devices \cite{novelHiGrad}. One of the key challenges in maintaining and improving the functionality of such devices at high fields is related to the control of field-induced breakdown \cite{rfBD} or vacuum arc generating events \cite{hopkins2020employing}.
As breakdown is routinely observed at fields that are much lower than those required for strong emission at flat clean surface \cite{DC_pulse_CERN_LES,rfBD}, the formation of sharp surface features which can locally enhance the applied field by factors of 50-100x is broadly assumed \cite{descoeudres2009dc, simakov2018advances}. Indeed, the development of breakdown from sharp emitter geometries is now well understood to follow from thermal runaway following from strong field emission currents \cite{nordlund2012defect}. 
The process of partial melting followed by plasma/vacuum arc formation have been desribed via a multiphysics and multiscale modeling approaches \cite{kyritsakis2018thermal}. The presence of a sharp emitter tip as has proved to be an essential trigger of breakdown in most of these studies \cite{eimre2015application, jansson2020growth}. 

However, to date, unraveling the precise mechanisms governing the formation of such sharp field-enhancing precursors has remained elusive, as they are destroyed in the breakdown process. Therefore, understanding the physical mechanisms through which the material surface couples and dynamically evolves under experimentally relevant field to form breakdown \textit{precursors} is paramount to design ultra-high gradient systems including the next generation of compact accelerators. Several hypotheses have been proposed to explain the formation of such precursors \cite{pohjonen2011dislocation, Yinon_stochastic,nordlund2012defect}. These hypotheses have been guided by the notion that real electrode deviate significantly from ideal materials. Material imperfections, such as point defects, voids, dislocations, slip extrusions, preexisting surface features, are thought to interact with electric fields, resulting in localized field-enhancement factors.  For example, the role of of dislocation mediated plasticity \cite{pohjonen2011dislocation} has been explored to address the formation of possible stress-concentrators through slips interacting with free metal surfaces. However, both fully-atomistic \cite{bagchi2022atomistic} and hybrid continuum-atomistic \cite{pohjonen2011dislocation} approaches suggest that fields on the order tens of GV/m (i.e. 2 orders of magnitude larger than typical experimental breakdown fields) are required to trigger substantial defect activity, unless these processes are assisted by other driving forces (e.g., thermo-elastic stresses caused by Joule heating \cite{laurent2011experimental}). As of now, these models remain difficult to directly validate experimentally, so the precise mechanisms leading to the formation of breakdown precursors at electric fields on the order of hundreds of MV/m remains poorly understood.


In this paper, we introduce a surface evolution model driven by surface tension and applied electric fields.
Our findings underscore the interplay between these forces: while surface tension tends to diminish local curvature, the electric field drives mass transport towards regions with heightened field strengths. This dynamic leads to the formation of sharper protrusions, further intensifying local fields and ultimately resulting in a tip-growth instability and potentially in breakdown.
Our investigation reveals a crucial aspect: the critical fields for instability exhibit a rapid increase with the growing aspect ratio of the initial surface morphology. For instance, in the case of a typical electro-polished copper surface featuring an average wavelength roughness of $4 \mu$m, our numerical modeling suggests that 
diffusive surface evolution can indeed lead to the formation of breakdown precursors at experimentally-relevant fields. Our investigation also reveals a crucial aspect: the critical fields for instability exhibit a rapid decrease with the growing aspect ratio of the initial surface morphology. The range spans from approximately 250 MV/m for higher aspect ratio profiles to 500 MV/m for nominally flat initial surface profiles.


The paper is organized as follows: first in Sec. \ref{sec:model} we describe the mesoscale finite element model of morphological evolution of surfaces under electric fields. We then analyze the subcritical and supercritical regimes and identify fields associated with unstable growth of surface perturbations into sharp protrusions in Sec. \ref{sec:results}. 
Finally, Section \ref{sec:discussion} puts these results in relation with previous experimental findings and discusses possible extensions of the model.


\begin{figure*}[hbt!]
\centering
\includegraphics[height=4in,width=1.\linewidth,angle=0]{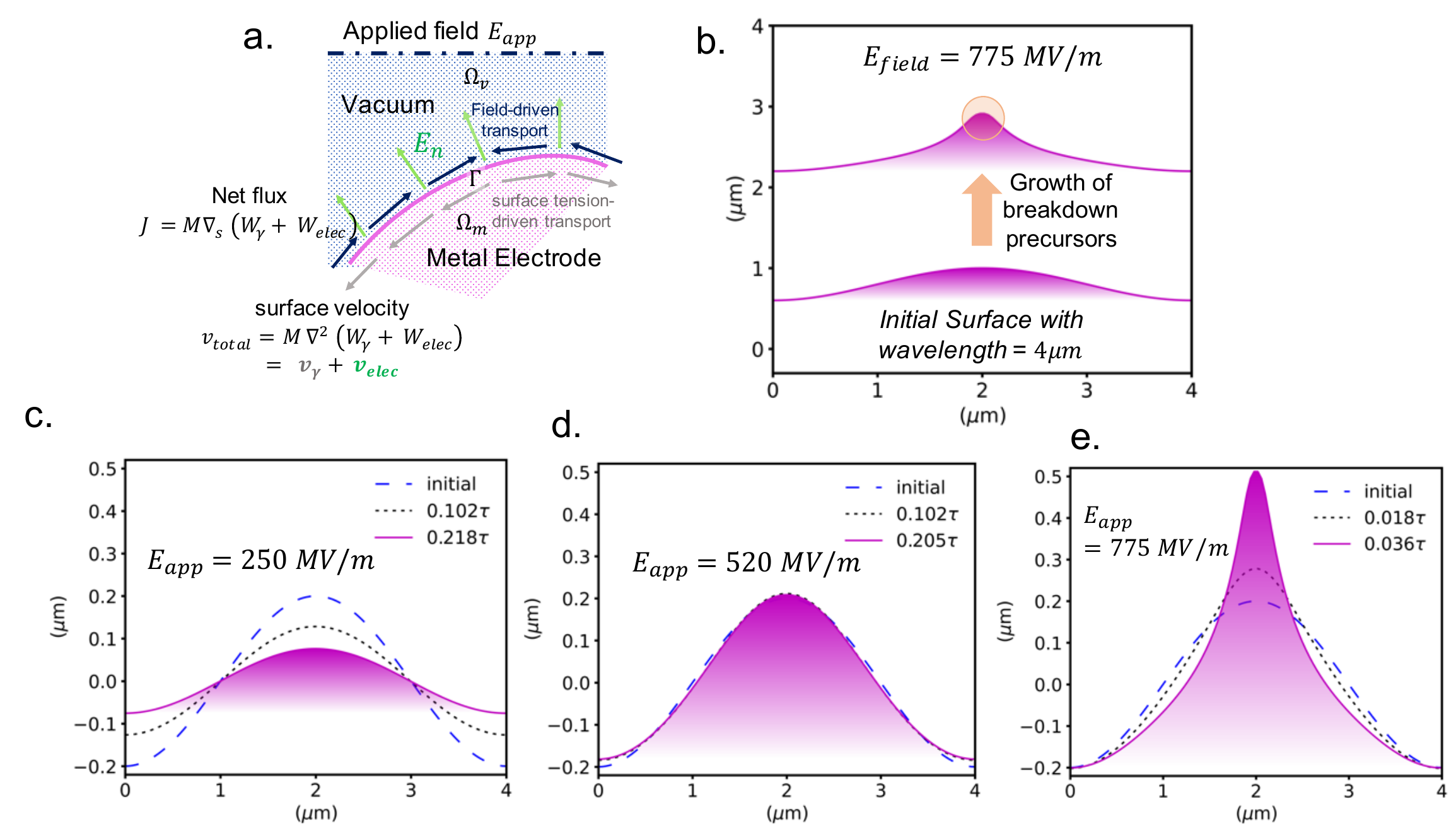}
\caption{\textbf{Curvature-driven growth of electrode surfaces under fields:} \textbf{a.} A metal electrode (magenta) is exposed to an applied electric field imposed at some distance in the vacuum (blue). 
Surface tension and electrostatic energies drive surface diffusion fluxes which leads to the evolution of the surface. 
\textbf{b.} Breakdown precursor formation at a nominally flat electro-polished copper surface perturbed at a wavelength ($\lambda$) $4 \mu$m \cite{electropolish_copper_surface} with a low aspect ratio ($A/\lambda \sim 0.1$) under $E_{field}=775$ MV/m. \textbf{c-e} illustrates the morphological evolution of a sinusoidal perturbation of aspect ratio 0.01 under various dynamical regimes. Under a subcritical field of 250 MV/m, the perturbation decays back toward a flat surface (\textbf{c}) Under a critical field of 520 MV/m, the surface perturbation remains stable. (\textbf{d}) Under a super-critical field (775 MV/m), a sharp tip quickly grows (\textbf{e}).}
\label{fig:Model setup and critical regime}
\end{figure*}

\section{Mesoscale curvature-driven growth model}\label{sec:model}
\textit{Surface kinetics}-- Our computational model follows conventional descent dynamics where 
the system evolves to minimize its free energy. 
Such models where first put forth by Herring \cite{herring1950effect, herring1999surface} and Mullins \cite{mullins1959flattening}, and have since proven successful at predicting several surface and interface evolution phenomena \cite{suo_microscopic_motions} e.g., grain boundary grooving \cite{suo_fem_1}, electromigration-driven void collapse \cite{Bower_Suo_FEM_electromirgation} in electronic interconnects, diffusive surface crack nucleation \cite{Srolovitz_nonlinear} and instability in stressed solids \cite{srolovitz1989stability} etc. In our setting, atomic migration is assumed to occur only at the free surface and to be governed by a driving force that is given by the surface-gradient of free energy density (\textit{per-unit volume}, $g$) ($\nabla_s g=\frac{\partial g}{\partial s}$). 


Depending on its wavelength ($\lambda$), a small-amplitude perturbation on the surface can either decay or grow based on competition between effective contribution to free-energy density from surface tension ($W_{surf}$--which tends to stabilize flat surfaces) and any external driving mechanisms ($W_{ext}$ -- potentially leading to growth of curvature). We further assume the flux of mass ($j_s$) to be proportional to the driving forces (Fig. \ref{fig:Model setup and critical regime}), i.e., 
\begin{align}{\label{eq:surface_diffusion}}
    j_s &= -M \nabla_s g \quad \textrm{in} \quad \Gamma\\
    &= -\frac{D \delta \Omega_{atom}}{k_BT} \nabla_s g \\
    &= -\frac{D \delta \Omega_{atom}}{k_BT} \nabla_s (W_{surf}+ W_{ext}).
\end{align}

Here, $M$ is the mobility of surface atoms which is related to surface self-diffusivity ($D=D_0e^{-E_b/k_bT}$ with $D_0 = 3.615^2/2\times 10^{12} \: \AA^2/s$ \cite{butrymowicz1973diffusion}) by $M = \frac{D \delta \Omega_{atom}}{k_BT}$, where $k_B$ is the Boltzmann constant, $T$  the temperature, $\Omega_{atom}$ the volume per atom, and $\delta$ the thickness of the diffusive layer. The surface free-energy contribution $W_{surf}$ is taken to be proportional to the surface curvature, i.e., $W_{surf}=\gamma \kappa $,
where $\kappa$ is the surface curvature
\begin{align}{\label{eq:surface_diffusion}}
    \kappa &= -\frac{\frac{\partial^2 h}{\partial x^2}} {(1 + {\frac{\partial h}{\partial x}}^2)^{3/2}},
\end{align}
 $\gamma$ is the surface tension, and $h(x)$ is the surface height at position $x$. In the following, $W_{ext}$ account for the coupling of the system with an externally applied electric field.

Invoking conservation of mass, a continuity equation for the flux leads to a the evolution equation
\begin{align}{\label{eq:mass_conservation}}
    v_y + \nabla_x . j_s &= 0 \quad \textrm{in} \quad \Gamma,
\end{align}
where $v_y$ is the local vertical velocity of the surface such that for a surface profile $h(x)$,

\begin{align}{\label{eq:velocity_surface_profile}}
    \frac{\partial h(x)}{\partial t} &= v_y,
\end{align}


\textit{Electrostatics}--Consider a domain as shown in Fig. \ref{fig:Model setup and critical regime}a which has a vacuum ($\Omega_v$) and an electrode material component ($\Omega_m$) which is assumed to be a perfect conductor. We obtain the electrostatic potential $\phi$ and the electric field $E=-\nabla \phi$)
through the solution of the Poisson equation :
\begin{align}{\label{eq:poisson}}
    -\Delta \phi &= 0 \quad \textrm{in} \quad \Omega_v \\
    \phi &= 0 \quad \textrm{on} \quad \partial \Omega_D \equiv \partial {\Omega_v}^{bot} \equiv \partial {\Omega_v} \cap \partial{\Omega_m} \equiv \Gamma\\
    -\nabla \phi.n &= E_{applied} \quad \textrm{on} \quad \partial \Omega_N \equiv \partial {\Omega_v}^{top}
\end{align}
with $E_{applied}$ the macroscopically-applied field. The electrostatic contribution to the free energy density at the surface is then obtained as $W_{elec}=\frac{1}{2}\epsilon {E_{n}}^2$, where $\epsilon$ is the vacuum permittivity, and $E_n$ as shown in Fig. \ref{fig:Model setup and critical regime}a is normal field acting on the surface $\Gamma$.
\\
\textit{Finite element formulation}--We solve the governing equations of surface kinetics using a mixed finite element formulation in two dimension. 
For test functions $u \in U$ and $w \in W$ defined over the surface manifold ($\Gamma$), we obtain a the variational weak form by integrating by parts:

\begin{align}{\label{eq:fem_surf_kinetics}}
    \int_{\Gamma}j_s u ds &= \int_{\Gamma}M g\nabla_s u ds \quad \textrm{,}
    \int_{\Gamma}v_y wdx &= \int_{\Gamma}j_s\nabla_xwdx 
\end{align}

\begin{figure*}[hbt!]
\centering
\includegraphics[height=4in,width=.8\linewidth,angle=0]{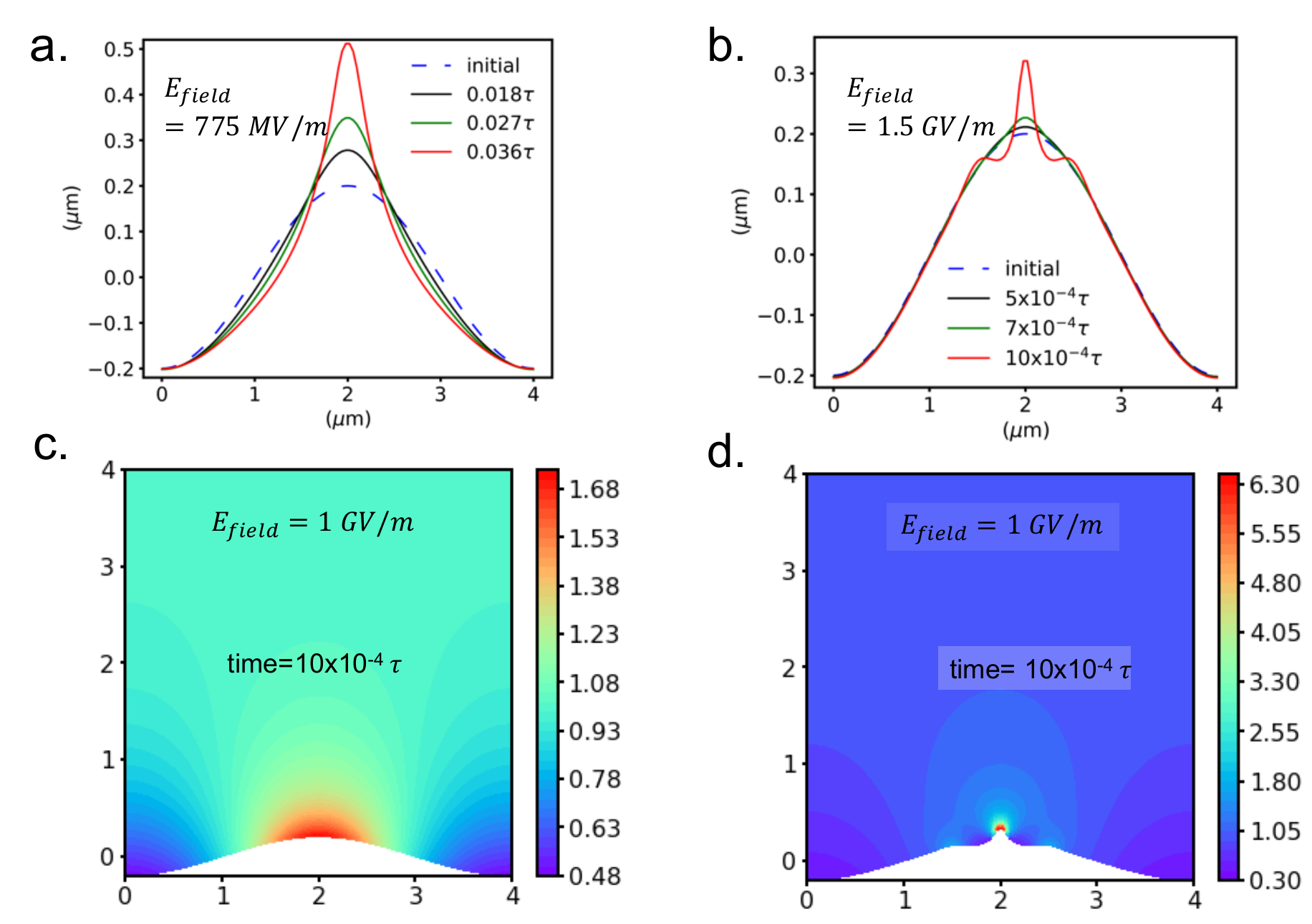}
\caption{\textbf{Breakdown precursor formation under strong e-fields:} At 775 MV/m (\textbf{a.}) and 1.5 GV/m (\textbf{b.}) Corresponding spatial distribution of the local electric field magnitude under an applied field of 1GV/m at time $t=10\times 10^{-4} \tau$. \textbf{c.}(\textit{case} \textbf{a.}) and \textbf{d.}(\textit{case} \textbf{b}).}
\label{fig:Effect of electric field}
\end{figure*}

The discretized solution space of Eq. \ref{eq:fem_surf_kinetics} is a mixed space $(j^h_s,v^h_n) \in (U,W)$. For both surface kinetics and electrostatics, we use Lagrangian elements of order 1. We currently model the surface as a 1D manifold ($\Gamma^h$) mesh. The 2D vacuum domain (${\Omega^h}_v$) for electrostatics is discretized with triangular elements using a finer mesh closer to the vacuum/material interface $\Gamma$, matching the nodes of the surface mesh. We use an explicit forward Euler time-integration scheme ($h(x, t+\Delta t) = h(x,t) + \Delta t v_y$) with self-adaptive timestepping depending on a maximum allowable surface displacement criteria to maintain numerical stability. 


In a staggered solution approach,  the electrostatics (Eq. \ref{eq:poisson}) is first solved  to evaluate $W_{elec}$ on the coinciding nodes of the meshes of the domains $\Omega_v$ and $\Gamma$, the surface profile $h(x)$
is then evolved using a small timesteps $\Delta t$. As the electrostatic solution changes slowly on the timescale 
$\Delta t$ required for numerical stability of the integration of the surface kinetic equations, 
the electrostatic solution is updated less frequently. A frequency ratio of 100-200 was found to give good results. 

In the resulting model, termed \textit{SurFE-XD} \textit{Surface curvature-driven Finite Elements model for Diffusion under eXtreme conditions}, the variational forms are specified in the FEniCS workflow \cite{solvingPDEfenics} in the high-level Python-based Unified Form Language (UFL) . These forms are then automatically compiled \cite{varcompiler} and executed through high-performance computational kernels using the finite element library DOLFIN \cite{dolfin}. 

\begin{figure*}[hbt!]
\centering
\includegraphics[height=5.5in,width=.8\linewidth,angle=0]{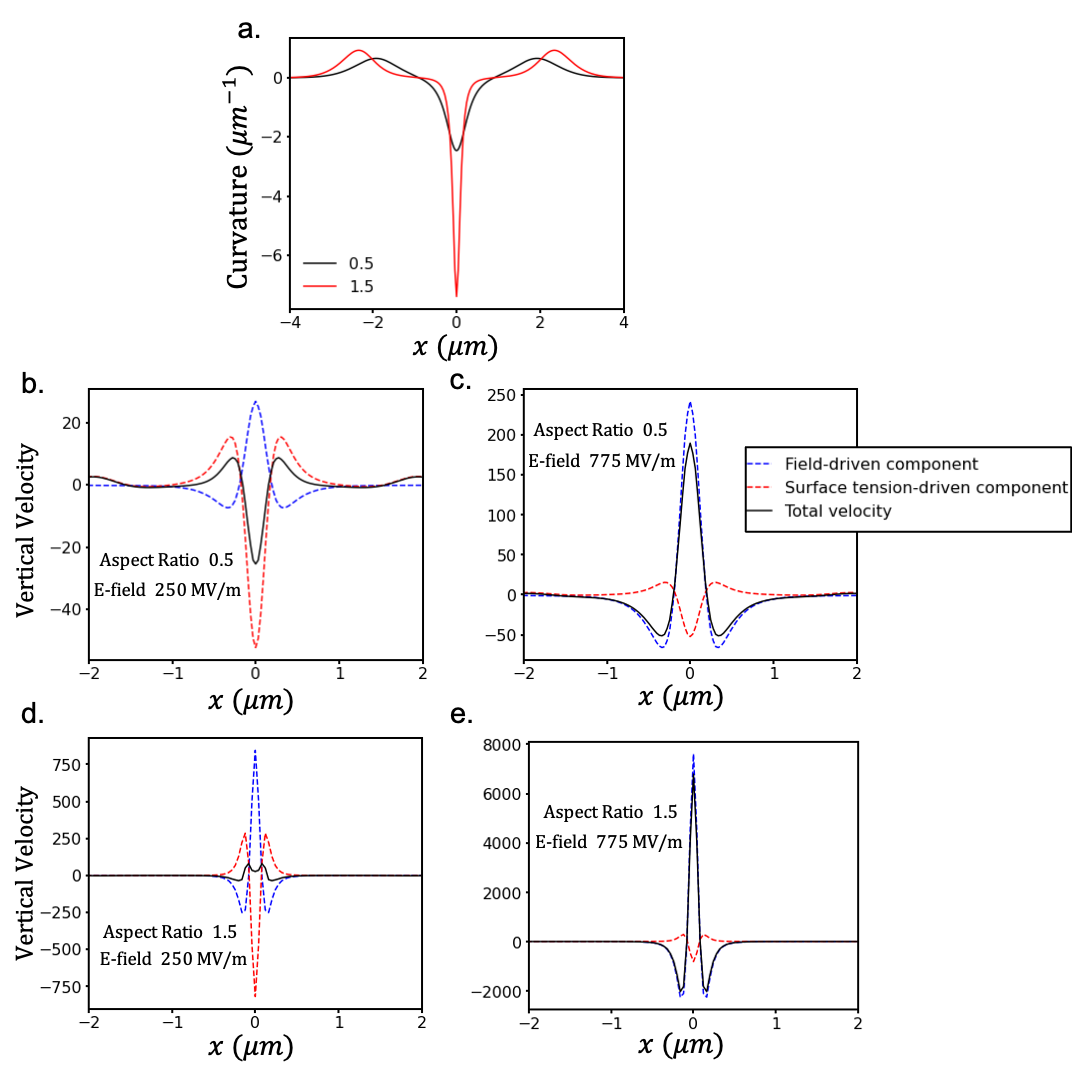}
\caption{\textbf{Effect of the initial aspect ratio on the relative contributions of surface tension and electric fields.} \textbf{a}. Curvature ($\kappa (x)$) of Gaussian surface profiles with aspect ratio 0.5 
and 1.5. Middle panel (\textbf{b,c}) with aspect ratio 0.5 and lower panel (\textbf{d,e}) with 1.5 show spatial variation of individual and total velocity contributions from electric field and surface tension driven forces under fields 250 MV/m (\textbf{b,d}) and 775 MV/m at $\tau=0$ (\textbf{c,e}).} 
\label{fig:velocity contribution}
\end{figure*}

\section{Results}\label{sec:results}

First, we demonstrate and verify our simulation results for a periodic sinusoidal profile with a low aspect ratio (amplitude/wavelength $\sim$ 0.1). Based on experimental measurements of electropolished copper surfaces \cite{electropolish_copper_surface}, we choose a wavelength of $4 \mu$m to present the results in terms of absolute physical units. We note that our continuum FEM framework only depends on relative geometry and is therefore scale-invariant, and the results can be rescaled for different wavelength, as we show in the following. 

In Fig. \ref{fig:Model setup and critical regime}, a critical field of 520 MV/m is identified below which surface tension dominates, leading to the gradual decrease in amplitude of the initial surface profile. 
Above the critical field, the electric field drives mass transport towards the tip, leading to the emergence of a sharp surface feature that has the characteristics of an efficient breakdown precursor (Fig. \ref{fig:Model setup and critical regime}e). 

At low aspect ratio, one can compare with growth/decay rates obtained from linear-stability analysis with an initial profile $h_0(x)=A_0 \sin kx, k=\frac{2 \pi}{\lambda}$. The time evolution of the surface amplitude is the given by $A(t)=A_0e^{\alpha t}$, where the rate $\alpha$ determines growth or decay under a given applied E-field ($E_{app}$) \cite{srolovitz1989stability}. One obtains $\alpha \propto (\frac{\epsilon E_{app}^2}{\gamma k}-1)$ leading to decay ($\alpha<0$) for $E_{app}<\sqrt{2\pi\gamma /\lambda\epsilon}$ and growth when $E_{app}>\sqrt{2\pi\gamma /\lambda\epsilon}$. Therefore, for a copper surface with $\lambda=4 \mu$ m, $\gamma=1.5$ J/$m^2$ \cite{kumykov2017surface} and $\epsilon=8.85\times10^{-12}$ F/m for vacuum, the critical field is predicted to be on around 520 MV/m. 

It is important to note that the critical field is inversely proportional to the wavelength $\lambda$, and so could be higher or lower depending on the physical scale of the surface roughness.
Our simulation results are in close agreement with the predictions of this linear stability analysis. One of the key limitations of this approximation is that is it only valid in the limit of small aspect ratios, i.e., $A_0/\lambda <<1$. Outside of this regime, explicit numerical solutions like that provided by (\textit{SurFE-XD}) is essential. 

In order to understand the effect of strong electric fields on the growth dynamics of nominally flat ($A/\lambda \sim 0.1$) copper surfaces, we analyze their evolution under a near-critical (775 MV/m) and stronger supercritcal (1.5 GV/M) regimes. Fig. \ref{fig:Effect of electric field}a shows that, at slightly super-critical fields, precursor formation occurs by long-range mass transport towards the hillock region. As the field are further increased into the super-critical regime,  the growth becomes much more localized, leading to the formation of sharp emitter (Fig. \ref{fig:Effect of electric field}b). The growth speed of the initial perturbation is also observed to be strongly dependent on the applied field. At 775 MV/m, the timescale required for growth up to an amplitude of 0.5 $\mu$m in the orders of $\sim 10^{-3}\tau$ in comparison with a timescales of about $\sim 3 \times 10^{-2}\tau$ at 1.5 GV/m.

Fig. \ref{fig:Effect of electric field} c-d shows that the electric field localizes at the hillock regions while it is being partially shielded at the trough or valley regions.  In the near critical regime (e.g., 775 MV/m), the contribution of surface tension towards mass transport is of similar order with that of the electric field. Hence, while local enhancement at the hillock region drives the curvature to grow, active mass transport 
from the valley towards the hillock lead to long-range mass transport. On the other hand, at 1.5 GV/m, the electrostatic driving force completely overwhelms surface tension in the hillock, leading to a localized and rapid growth of a very sharp tip. The rapid nonlinear growth towards instability is accelerated by local field enhancement in highly-curved regions. 

To further probe the effect of local geometry on field induced breakdown precursor formation, we model the evolution of isolated Gaussian-shaped features with aspect ratios ranging from very low (0.1) to high (2.0). For a similar wavelength of 4 $\mu$m, Fig. \ref{fig:velocity contribution}a depicts the sharp magnification of curvature at the tip of the Gaussian. This leads to a drastic change in the surface evolution velocity, leading to a surface evolution that is more localized, larger in magnitude, and faster, as the aspect ratio becomes larger.

For a fixed width of $4 \mu$m, the individual contributions of surface tension ($v_{\gamma}$) and electrostatics ($v_{elec}$) again act in opposite directions, as shown in Figs \ref{fig:velocity contribution}b-e. 
Under a field of 250 MV/m, for a surface feature with a modest aspect-ratio (0.5), (Fig. \ref{fig:velocity contribution}b) surface tension slightly dominates over the electrostatic driving force at the tip location ($x=0$), leading to the decay of the initial protrusion. This trend is completely reversed at a field of 775 MV/m for the same aspect ratio 0.5 (Fig. \ref{fig:velocity contribution}c); here the electric-field-induced velocity drives the growth of the tip resulting in the rapid growth of of sharp feature that could act as a breakdown precursor.

\begin{figure}[t]
\centering
\includegraphics[height=2.5in,width=3in,angle=0]{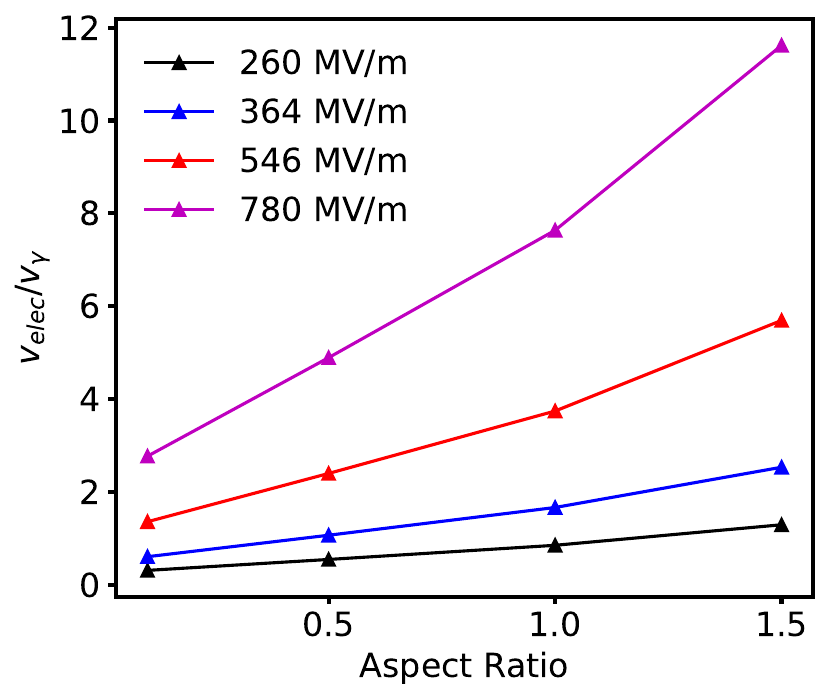}
\caption{\textbf{Ratio of the vertical surface velocity contributions stemming from electrostatics and surface tension} at the location of the tip as a function of aspect ratio for a Gaussian protrusion of $4 \mu$m in width.} 
\label{fig:ratio}
\end{figure}

At even higher aspect ratios (1.5, c.f. Fig \ref{fig:velocity contribution}d)
the local field enhancement at highly curved regions magnifies the electro-static driving forces for diffusion, inducing tip growth at fields that were previously sub-critical for smaller protrusions. E.g., at field of 250 MV/m, the electrostatics contributions slightly exceed the restoring tendency of the surface tension, resulting in a net positive velocity at tip. This is in contrast with the much higher linear stability estimation of 520 MV/m reported above. We note that two effects are at play: first, field enhancement is more efficient at isolated protrusions due to the lack of interactions between tips (however note that in our simulations, periodic boundary conditions remain in the lateral direction so tip-tip interactions are not totally absent, but they are smaller in magnitude);  second, non-linear interactions between modes invalidate the predictions obtained in the linear regime.

\begin{figure*}[hbt!]
\centering
\includegraphics[height=4.8in,width=5in,angle=0]{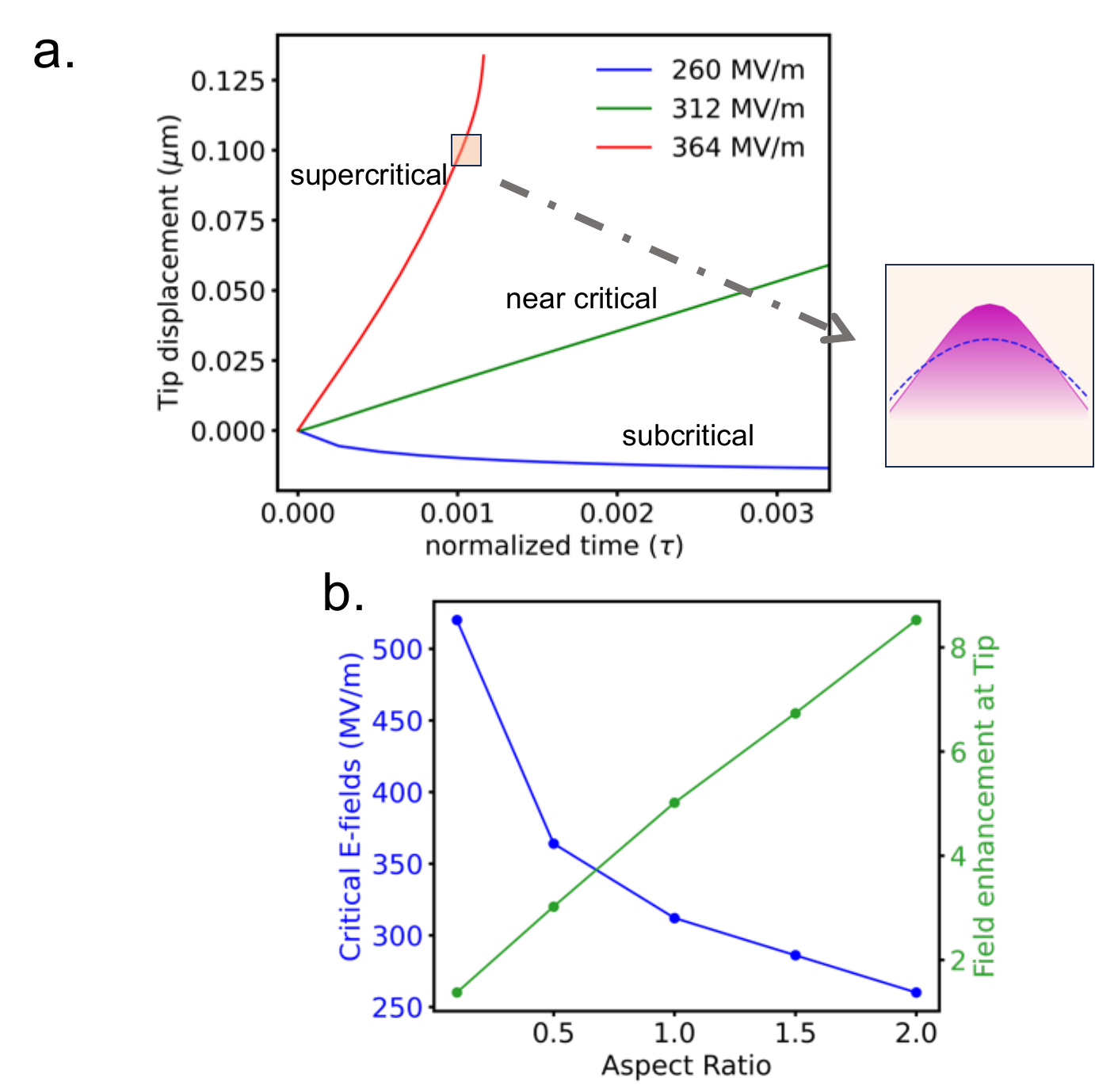}
\caption{\textbf{Time evolution of precursor growth for identifying critical electric fields:} \textbf{a.} For an aspect-ratio of 1.0, subcritical (260 MV/m), near-critical (312 MV/m) and supercritical (364 MV/m) regimes of tip-displacement evolution. \textbf{Inset} Sharpening of tip (magenta fill) at supercrtical (364 MV/m) compared to initial tip geometry in blue dashed line. \textbf{b.} Critical e-fields (blue) decrease with increasing aspect ratio while field enhancement at the tip (green) linearly increases.} 
\label{fig:time-evol and critical fields}
\end{figure*}

The respective contributions of surface tension and electric fields is further analyzed in Fig. \ref{fig:ratio}, 
which presents the ratio of the different velocity contributions ($\frac{v_{elec}}{v_{\gamma}}$ at the tip of the Gaussian feature. Here, the sub-critical and super-critical regimes correspond to $v_{elec}/v_{\gamma}$ <1 and $v_{elec}/v_{\gamma}$ > 1, respectively. The results show that the velocity ratio grows 
 super-linearly as the aspect-ratio ($A/\lambda$) and that the effect becomes even more prominent when the applied fields become stronger. At a low aspect-ratio of 0.1, we observe $v_{elec}/v_{\gamma} \sim 1$ at 546 MV/m, consistent with linear stability predictions. However, with larger aspect-ratio (e.g., 1.5), the velocity ratio
 is around 5 at this same field. The velocity comparison due to fields of 250 MV/m and 775 MV/m in Fig \ref{fig:velocity contribution} show almost an order of magnitude difference when aspect-ratio is even larger (1.5), demonstrating that the balance quickly shifts in favor of electro-static contributions as the aspect ratio increases, leading to tip-growth instabilities at progressively lower fields. As can be seen in Fig. \ref{fig:time-evol and critical fields}b, the crtical field for the tip to grow can reduce from 520 MV/m to ~250 MV/m as the aspect-ratio increases from nominal to a higher value of 2.0. This is in line with the 4-fold increase in field-enhancement near tip (Fig. \ref{fig:time-evol and critical fields}b). 
 
 

\section{Discussion}\label{sec:discussion}

A key factor that should be quantified in order to assess the efficiency of this mechanism is the timescale over which the growth of such surface features can be expected to occur. 
The time-unit in the simulation $\tau$ is estimated as:
\begin{equation}
 \tau=\frac{1}{\gamma M}
\end{equation}

Considering Cu parameters $D=D_0e^{-E_b/k_bT}$ with $D_0 = 3.615^2/2\times 10^{12} \: \AA^2/s$ \cite{butrymowicz1973diffusion}; $\delta=3.615 \: \AA$ and $\Omega=11.81 \: \AA^3$ \cite{simon1992properties}), $\tau$ can range from 475 seconds to 13 days. Such a wide range in physical timescales is due to the exponential dependence of surface ad-atom diffusivity on energy barrier ($E_b$) which is changes considerably with change in local environment; for instance, $E_b$ could range from 0.1-0.15 eV for (111), 0.25-0.30 eV for (110), and 0.38-0.69 eV for (100) \cite{Eb_1,Eb_2, Eb_3, Eb_4}.
It is difficult to lump atomistic details on a single effective diffusive property without more sophisticated multiscale investigation which is currently beyond the scope of current study.

Nonetheless, the simulations have shown that breakdown precursors can be expected to spontaneously form through surface transport at fields in the range of 200 to 500 MV/m, depending on the initial roughness of the surface. This mechanism can therefore potentially explain the spontaneous formation of breakdown-inducing features in application-relevant conditions. Indeed, the conditions investigated here are consistent with experiments carried on for breakdown with DC pulses with Copper electrodes \cite{DC_pulse_CERN_LES}. These experiments have shown features: i) the spatial location of the breakdown events is highly correlated, with a high fraction of craters overlapping, and ii) the breakdown events occurred preferentially close to the edge of the electrode. The first point is consistent with the observation that breakdown events leave craters and debris behind \cite{DC_pulse_CERN_LES, PhysRevB.81.184109}. As shown above, these surface features can couple efficiently with the electric field, potentially leading to fast tip-growth instabilities and to further breakdown events occurring in close proximity and in rapid succession, consistent with experimental results \cite{PhysRevAccelBeams.20.011007,PhysRevB.81.184109}. Second, the edge of the cathode will be exposed to additional macroscopic field gradients which can also drive surface diffusion. This effect has been observed in nanoscale experiments on gold surfaces where electric field gradients at the edge of the electrode led to enhanced formation of ridge and tip-like features \cite{AFM_electrodiffusion_edge_roughness}. Both these observations are consistent with the hypothesis that breakdown was there mediated by surface diffusion driven by electric field gradients. 

Finally, our simulations show that the mechanism investigated here can lead to very-fast runaway instabilities. Indeed, Fig. \ref{fig:time-evol and critical fields}a shows that tip growth proceeds in two stages: an initial stage where the tip grows roughly linearly in time, and a second runaway stage where exponential growth is observed. This suggests that directly observing these sharp precursors in experiments can be expected to be very challenging, as the onset of runaway growth occurs at relatively modest aspect ratios and then unfolds very rapidly, leading to the destruction of the tip during the breakdown process. During short-time scanning probe investigations of surface roughness of electrodes in vacuum under moderate electric fields, sharp features leading to high field-enhancement factors were rarely observed \cite{hopkins2017vacuum}. This is consistent with a mechanism where sharp breakdown precursors form only shortly before breakdown occurs.


While field-driven surface diffusion appears to be a viable mechanism to explain the formation of 
breakdown precursors, other phenomena could contribute. For example, theoretical treatments typically assume
clean surfaces free of contamination. Any contaminants could potentially affect surface evolution in complex fashion. Second, many high-field applications occur in high-frequency AC settings \cite{rfBD,simakov2018advances}. In this case, losses through Joule heating induce thermo-elastic stresses. These introduce an additional driving force for surface deformation \cite{Srolovitz_nonlinear} and can also 
lead to the formation of surface roughness through thermal fatigue \cite{laurent2011experimental}.
Other mechanisms could therefore "seed" the tip-growth process by introducing surface features, which would then 
grow and sharpen through the diffusive mechanism discussed here. The inclusion of thermo-elasticity in \textit{SurFE-XD} is currently underway and the results will be presented in a subsequent manuscript \cite{ryo_Shino_thermal_stress}.

\section{Conclusion}\label{sec:Conc}
In this study, the evolution of metallic surfaces exposed to strong electric fields was modeled using a 
surface diffusion model driven by surface tension and electro-static driving forces. 
The numerical results are consistent with the prediction of linear stability analysis in the regime small aspect-ratio perturbations on the surface. Non-linearities in the growth process start to emerge with high aspect-ratio surface features as well as under stronger electric fields. Our results show that exposure to electric fields in the range of 250-500 MV/m can trigger the spontaneous formation of sharp surface features that can act as precursors for breakdown. These fields are of the same order of magnitude as those experimentally observed to lead to breakdown.


\section*{Acknowledgments}
This work was supported by the Laboratory Directed Research and Development program of Los Alamos National Laboratory under project number 20230011DR. Los Alamos National Laboratory is operated by Triad National Security, LLC, for the National Nuclear Security Administration of U.S. Department of Energy (Contract No. 89233218CNA000001).

\bibliography{references}


%



\end{document}